\documentclass[aip,reprint,amsmath,amssymb,superscriptaddress]{revtex4-1}

\usepackage{graphicx}
\usepackage{graphics}
\usepackage{bbm}
\usepackage{amsmath,amsfonts,amssymb}
\usepackage{wrapfig}

\begin{document}
\title{Superconducting Resonators with Parasitic Electromagnetic Environments}
\author{J. M. Hornibrook}
\affiliation{ARC Centre of Excellence for Engineered Quantum Systems, School of Physics, The University of Sydney, Sydney, NSW 2006, Australia}
\affiliation{CSIRO Materials Science and Engineering, Lindfield, NSW 2070, Australia}
\author{E. E. Mitchell}
\affiliation{CSIRO Materials Science and Engineering, Lindfield, NSW 2070, Australia}
\author{D. J. Reilly$^*$}
\affiliation{ARC Centre of Excellence for Engineered Quantum Systems, School of Physics, The University of Sydney, Sydney, NSW 2006, Australia}


\begin{abstract}
Parasitic electromagnetic fields are shown to strongly suppress the quality ($Q$)-factor of superconducting coplanar waveguide resonators via non-local dissipation in the macroscopic environment. Numerical simulation and low temperature measurements demonstrate how this parasitic loss can be reduced, establishing a Lorentzian lineshape in the resonator frequency response and yielding a loaded $Q$-factor of 2.4 $\times$ $10^5$ for niobium devices on sapphire substrates. In addition, we report the dependence of the $Q$ and resonance frequency shift $\Delta f_0$ with input power and temperature in the limit where loss from two-level systems in the dielectric dominate. 
\end{abstract}

\maketitle

The viability of quantum information hardware based on condensed matter is dependent on isolating and protecting quantum systems from environments that lead to dissipation and uncontrolled evolution \cite{Zurek91,*Caldeira81}. On-chip microwave resonators are key components in quantum technology, enabling readout \cite{Wallraff04}, creating strong interaction between distant qubits \cite{DiCarlo09}, and providing a means of transporting a quantum state between different architectures, for instance, in the coherent coupling of superconducting qubits to spins \cite{Schuster10}. Understanding the mechanisms that lead to dissipation in resonators is thus of central importance in scaling-up quantum information processing and in the construction of supporting quantum technology such as single photon detectors \cite{Day03} and parametric amplifiers \cite{Tholen07}.

At low temperatures and frequencies below a few 100 GHz, intrinsic dissipation in superconducting devices is dominated by loss from dielectric materials \cite{OConnell08} and radiative processes that depend strongly on the electromagnetic environment \cite{Houck08,Paik11,Wenner11}. Recent work has also investigated loss from trapped Abrikosov vortices \cite{Song09} and stray infrared light \cite{Barends11}. In addition to these loss mechanisms, practical resonator circuits are always loaded by lossy external measurement and excitation circuitry. For device applications that require  $in$ $situ$ high quality ($Q$) factor resonators, loading from external circuitry should not lead to a further suppression in the $Q$-factor from its intrinsic limit. 

\begin{figure}
\includegraphics[scale=0.4]{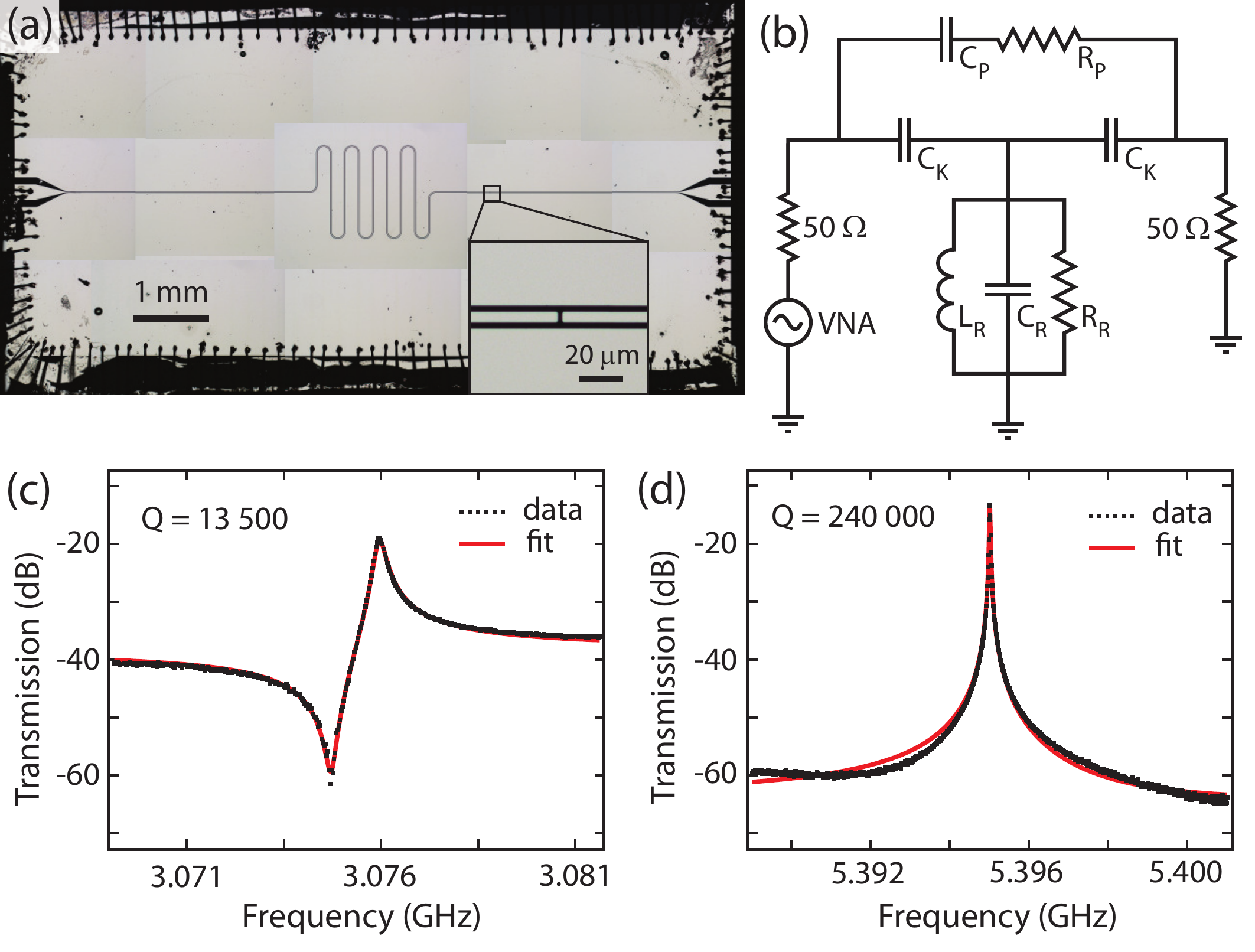}
\caption{\label{fig:photos}  \textbf{(a)} Photograph of a niobium CPW resonator on sapphire, inset shows 0.5 fF coupling capacitors. \textbf{(b)} Equivalent circuit model for the resonator and parasitic path ($C_P$) with dissipative element ($R_P$). \textbf{(c,d)} show transmission measurements with -90 dBm power after attenuation at a temperature of 10 mK. In \textbf{(c)}, a parallel dissipative path reduces the $Q$ and results in an asymmetric lineshape. In \textbf{(d)}, the parallel path is suppressed resulting in a higher $Q$ and  Lorentzian lineshape. }
\end{figure}

In this Letter, we investigate how a dissipative parasitic environment loads superconducting coplanar waveguide (CPW) resonators, strongly suppressing the $Q$-factor and leading to a Fano lineshape in the frequency  response of the resonator. We demonstrate via electromagnetic (EM) simulation and low temperature measurements, how this dissipative parasitic coupling can be reduced, restoring a Lorentzian line shape and yielding a $Q$-factor of 2.4 $\times 10^5$ for niobium devices on sapphire substrates. Having suppressed losses from parasitic coupling, we further report the dependence of the loaded $Q$ and resonance frequency with input power and temperature. In the low power regime, measurements are consistent with dissipation from two-level systems (TLS) associated with defects in the dielectric \cite{OConnell08, Macha10, Gao08}.

CPW half-wavelength ($\lambda/2$) resonators are patterned using optical lithography and argon ion-beam milling of 150 nm thick niobium films on r-cut sapphire substrates as shown in Fig. 1(a). Substrates are first cleaned using the ion-beam before sputter deposition of Nb at $3\times 10^{-7}$ mb. Niobium films yield critical current densities of $J_c \sim$ 14 MAcm$^{-2}$ and critical temperatures of $T_c \sim$ 8.3 K. The geometry of the CPW resonator comprises a 10 $\mu$m wide central track separated from ground on both sides by gaps, 4.6 $\mu$m wide, defining a characteristic impedance $Z_0$ = 50 $\Omega$. The resonator is under-coupled to highly filtered and attenuated input and output ports via two gaps in the central track that define 0.5 fF capacitors [Fig. 1(a), inset]. Numerous aluminium wire bonds are used to ground the Nb device to a PCB (Rogers dielectric RO6010) that is soldered into a light-tight copper enclosure and mounted at the mixing chamber stage of a dilution refrigerator (with base temperature 10 mK) after 60 dB of attenuation from room temperature. $Q$-factor measurements are made using a vector network analyser to determine S-parameters after cryogenic \cite{LowNoiseFactory} and room temperature amplification. 

A lumped element equivalent circuit of the resonator is shown in Fig. 1(b). In addition to the capacitance and inductance per unit length that define the resonator, the equivalent circuit accounts for a parallel circuit path arising from parasitic coupling between the ports and resonator. The addition of this parasitic path $C_P$, which is always present to some degree, leads to a Fano lineshape in the frequency response of the resonator, shown in Fig. 1(c) for a $\lambda/2$ = 20.8 mm device at base temperature. In our equivalent circuit, $C_{k}$ represents the combined parallel capacitance of the coupling ports and parasitic coupling from the central conductor. We extract a $Q$-factor of 13500 from this measurement by fitting [red line in Fig. 1(c)] to the functional form of a Fano resonance, $\sigma (\epsilon)=(\epsilon+q)^2 / (\epsilon^2+1)$, where $\sigma$ is the frequency dependent amplitude, $\epsilon$ is the frequency detuning and $q$ is a parameter that characterises the strength of the Fano lineshape. 
 
A Fano lineshape has been observed previously for CPW resonators \cite{Paik10, Sage11,Khalil12} and can be understood as arising from the coherent interaction between the single frequency mode of the resonator and a continuum of accessible modes from a parallel circuit path that has a flat frequency response \cite{Miroshnichenko10}. Importantly, we note that the presence of this parasitic parallel path, and its Fano lineshape signature, do not constitute a loss mechanism unless the parallel path is dissipative. We find however, that for practical circuit implementations, this parasitic path involves resistive losses from normal metals in the PCB and dissipation in dielectric materials well beyond the neighbourhood of the resonator. Below we show how this parasitic dissipation leads to a strong reduction in the loaded $Q$-factor. 

We first compare the frequency response of two very similar resonators, each bonded to different PCBs and sample enclosures. The results from these two devices are indicative of several other devices we have measured, including resonators fabricated on different substrates \cite{extras1}. Comparing the lineshape, Fig. 1(c) and (d) show that the frequency response is very sensitive to the details of the PCB and sample enclosure interconnects. For the Fano lineshape shown in Fig. 1(c), measurements were made in the sample mount shown in Fig. 2(a), while the Lorentzian lineshape data in Fig. 1(d) was taken using a sample mount designed to suppress parasitic coupling, shown in Fig. 2(b). In addition to the variation in lineshape between the two data sets, we note the 20-fold difference in $Q$-factor that results from different parasitic dissipation in the two sample mounts. Fits to the data [shown in red in Fig. 1(c) and (d)] yield an estimate of the strength of the parasitic coupling characterised by the Fano parameter $q$. For low parasitic coupling ($q \sim$  1000) the lineshape becomes Lorentzian, characteristic of a driven oscillator. 

\begin{figure}
\includegraphics[scale=0.4]{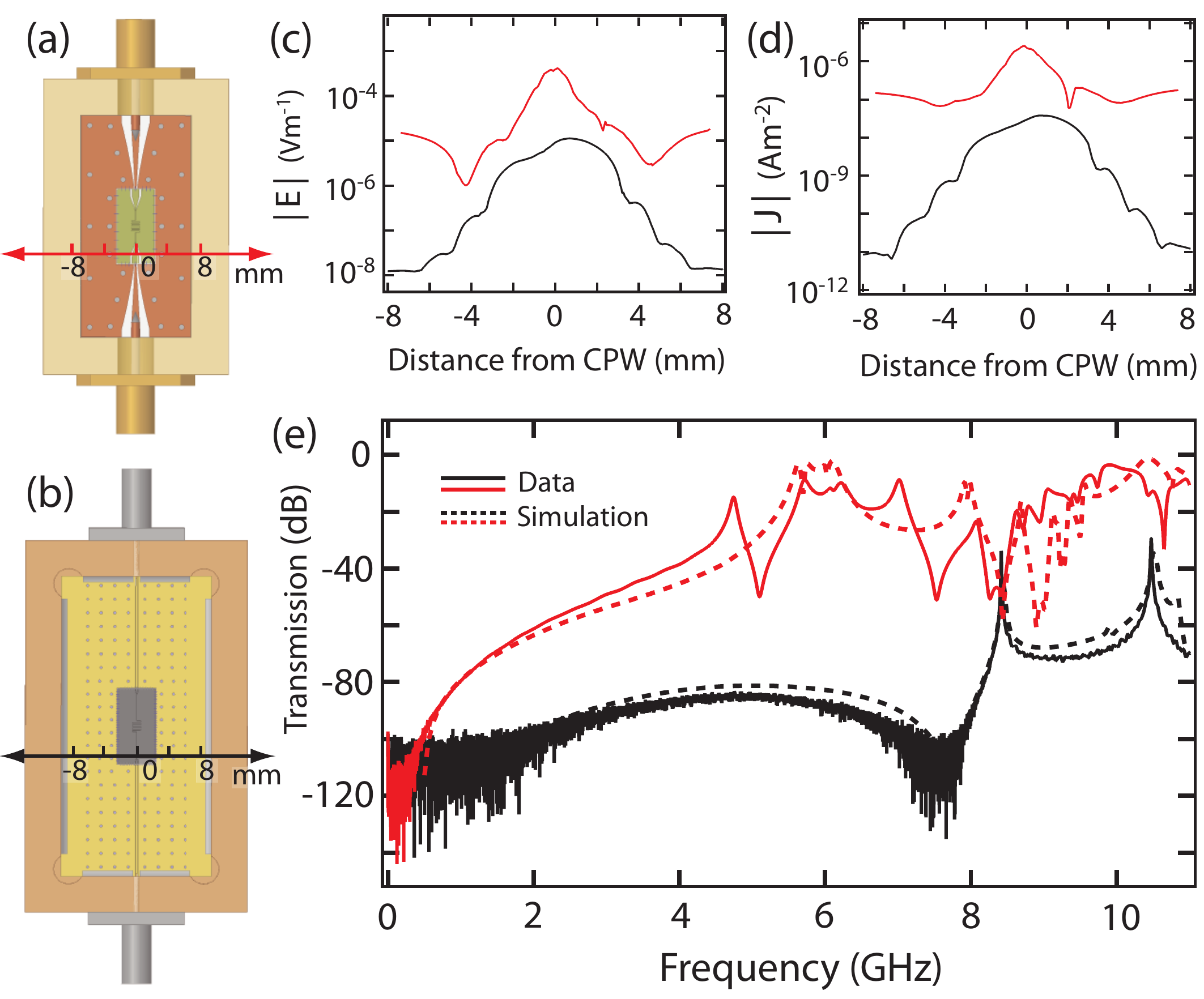}
\caption{\label{fig:crosstalk} \textbf{(a), (b)} Two different sample setups with different parasitic coupling used in measurements shown in Fig. 1(c) and 1(d), respectively. \textbf{(c)} Electric field $E$ in the PCB dielectric for the slices indicated by the axes in (a), (b) and \textbf{(d)} current density $J$ in the lower PCB ground plane along the same slice. \textbf{(e)} Comparison of simulated (dashed lines) and measured (solid lines) transmission response for setups in (a) and (b). Data taken at room temperature in the absence of a superconducting chip.}
\end{figure}

To better understand this parasitic dissipation, we simulate the environment of the resonator, PCB, and sample enclosure using a finite element 3D EM field solver \cite{Ansoft}. Taking a horizontal slice across the device, we compare the simulated electric field and current densities present for the two different setups shown in Fig. 2(a) [red] and (b) [black]. The magnitude of the E-field slice [Fig. 2(c)] is taken in the dielectric, 250 $\mu$m below the copper surface of the PCB and the current density is calculated at the lower PCB ground plane [Fig. 2(d)]. As is evident in the simulation, the parasitic field density has been significantly lowered for the sample mount arrangement shown in Fig. 2(b). To confirm that these EM simulations accurately capture the response of our system, we also compare the simulated response of the enclosure and PCB, (with no resonator chip) to transmission data measured for both configurations as shown in Fig. 2(e). We note that for the sample mount shown in Fig. 2(b) we have suppressed the parallel coupling between ports by $\sim$ 60 dB [see Fig. 2(e)].

\begin{figure}
\includegraphics[scale=0.4]{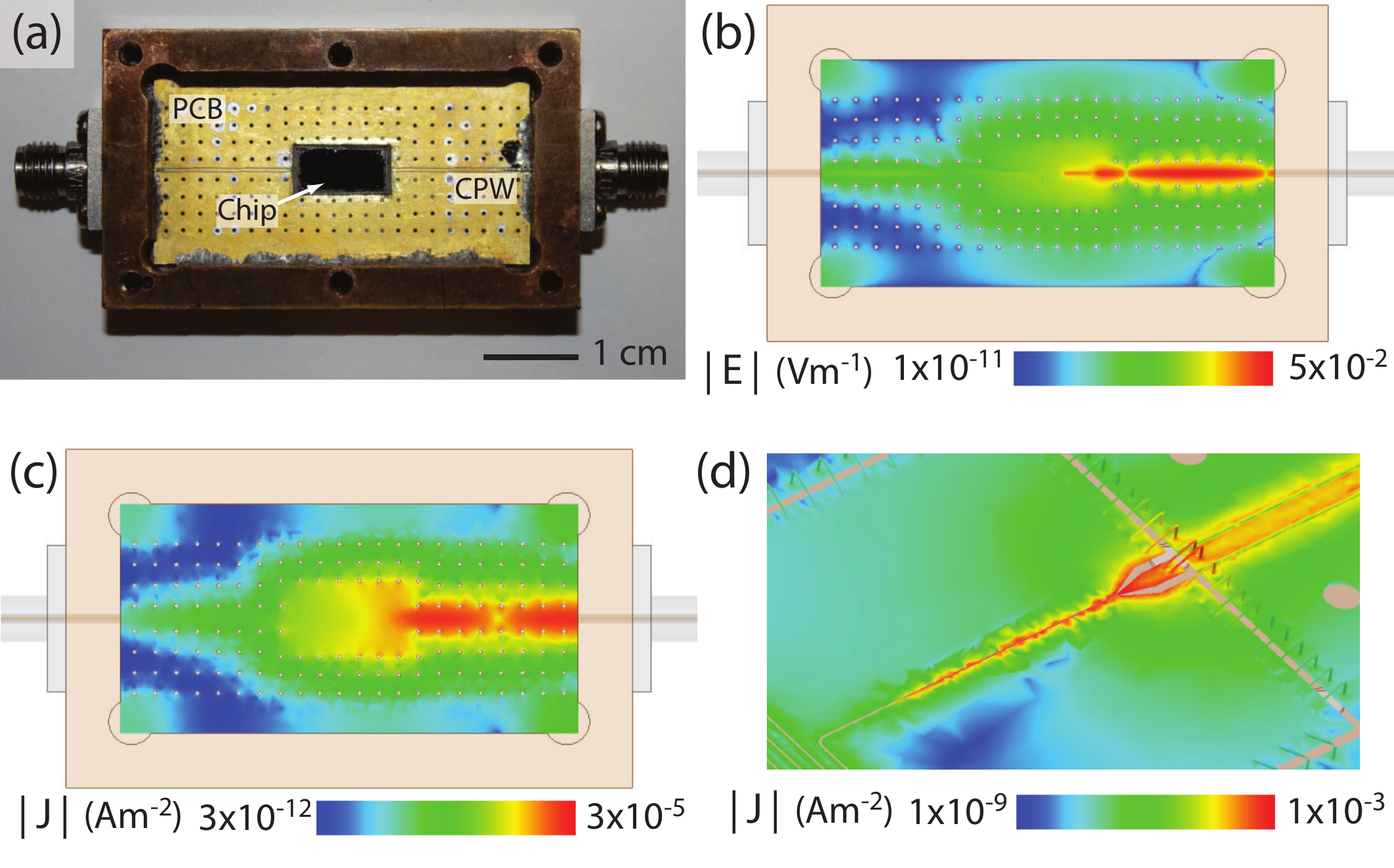}
\caption{\label{fig:test}  \textbf{(a)} The chip with CPW is wire bonded to the PCB that incorporates a coplanar line and via fencing between sample and SMA connectors. Simulations of the parasitic EM signal using Ansoft's HFSS at -90 dBm input powers for a frequency near $f_0$. \textbf{(b)} E-field magnitude in the PCB dielectric, \textbf{(c)} current density in the PCB ground plane, and \textbf{(d)} the current density in the chip, top PCB ground planes, and wire bonds.}
\end{figure}

The large electromagnetic cross-section of superconducting resonators makes them susceptible to parasitic effects on macroscopic scales. To visualise the extent of non-local EM fields, Fig. 3(b-d) shows further results of our EM simulations for the low parasitic setup shown schematically in Fig. 2(b) and as a photograph in Fig. 3(a). Even for this optimised setup, we find significant electric field and current density ``hot-spots" far from the resonator. These are largely associated with regions in which the geometry of the planar transmission line varies, despite a constant impedance, such as at the bondwire interface between the PCB and superconducting chip [see Fig. 3(d)]. We believe it is these regions of appreciable current and E-field density that constitute the parasitic environment of the resonator, producing dissipation in the PCB normal metal, dielectric, and bondwire interconnects. 

In moving from the sample mount design shown in Fig. 2(a) to the low dissipation mount shown in Fig. 2(b) we have reduced parasitic coupling by adding numerous tightly spaced vias to ground that strongly confine the E-field and current \cite{Colless12}. In addition, the design does not taper the transmission line \cite{Hornibrook12}, employing microwave launchers and CPW track widths of the smallest possible size. A further reduction in parasitic coupling was evident for well-matched transmission line geometry at the PCB-chip interface. Numerous wirebonds are used to reduce on-chip ground current density \cite{Barends11}.

Finally, having alleviated parasitic coupling as the dominant source of dissipation, we report the dependence of the loaded $Q$ on input power and temperature. It is now well established that defects or TLSs in the resonator dielectric lead to loss by absorbing microwave photons at low power and temperature \cite{OConnell08,Macha10,Gao08}. For the low dissipation setup, we observe an improvement in $Q$ with increasing input power [Fig. 4(a)] consistent with TLSs being driven into long-lived excited states that cannot absorb further microwave energy. At still higher powers a strong reduction in $Q$ and a distortion in the lineshape is evident [see Fig. 4(b)], consistent with a non-linear surface impedance from large-angle grain boundaries \cite{Abdo06, Oates02}. Further evidence that the loss is now dominated by TLSs is seen in the non-monotonic temperature dependence of the loaded $Q$ [Fig. 4(c)] and fractional frequency shift $\Delta f/f_0$ [Fig. 4(d)].  At temperatures below 100 mK the loaded $Q$ and $\Delta f/f_0$ exhibit a slight inflection that is presumably derived from the thermal population difference between the excited and ground states of the TLSs \cite{Lindstrom09}.
\begin{figure}[t]
\includegraphics[scale=0.4]{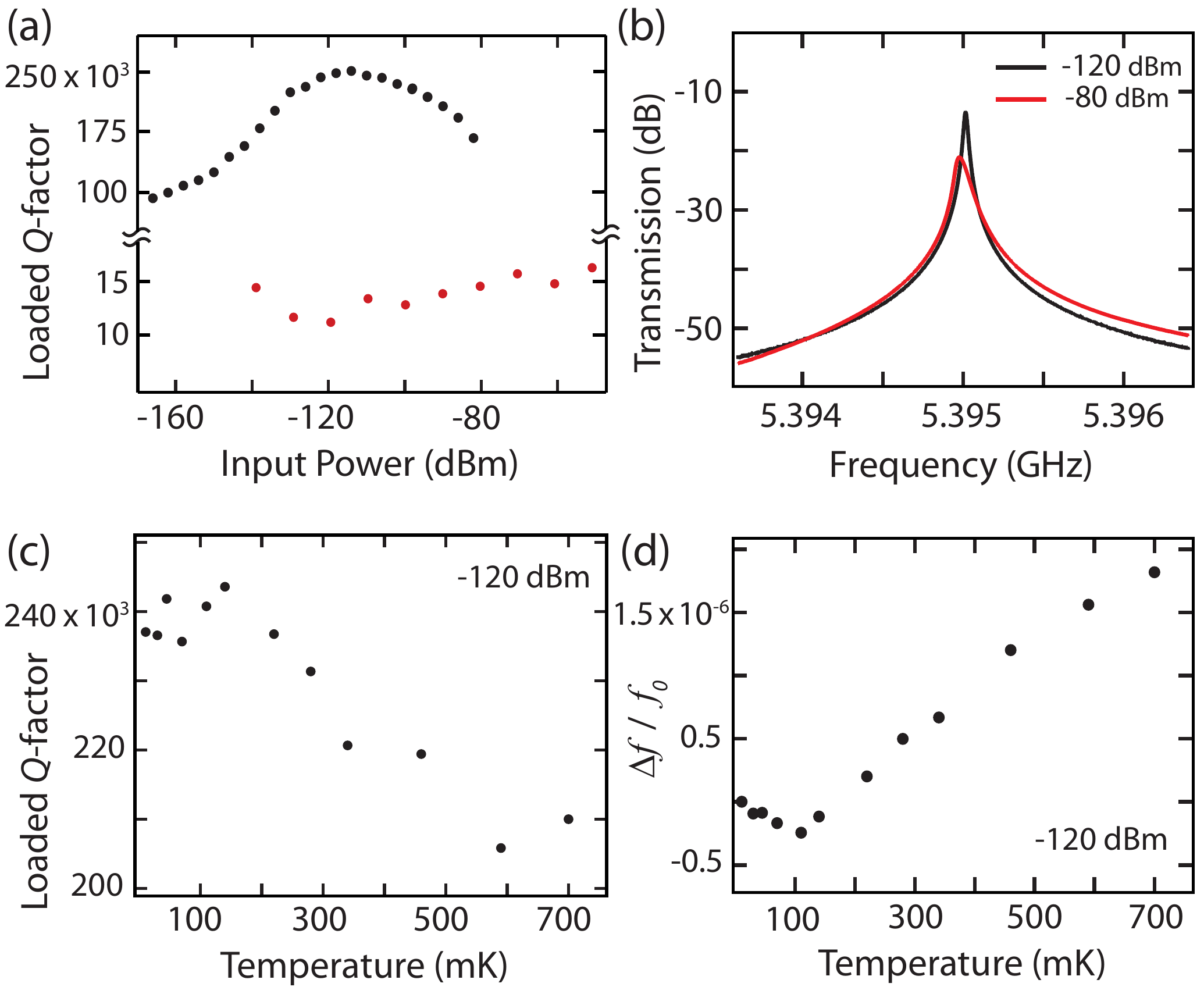}
\caption{\label{fig:sims} \textbf{(a)} Power dependence of the loaded $Q$ for the two setups with different parasitic dissipation where black points are from the setup in Fig. 2(a) and red points from setup in Fig. 2(b). Data taken at 10 mK. \textbf{(b)} Nonlinear response of the resonator at high input power. \textbf{(c)} shows the temperature dependence of the loaded $Q$ and \textbf{(d)} the resonance frequency fractional shift $\Delta f/f_0$ attributed to the presence of TLSs in dielectric environment of the resonator.}
\end{figure}

The presence of macroscale parasitic channels, as investigated here, lead to an unwanted coupling between the resonator and its dissipative environment. For complex device architectures that require many ports and planar microwave feed-lines, parasitic modes that inadvertently couple energy far from the resonator present a key technical challenge for low loss quantum circuits. We anticipate that the mitigation of parasitic dissipation using the methods reported here will be of interest for the design of scaled-up quantum hardware.

We acknowledge C. J. Lewis and J. I. Colless for technical assistance and thank J. Martinis, A. Houck, J. Aumentado and T. Duty for useful discussions. This research was supported by the IARPA/MQCO program through the U. S. Army Research Office and the Australian Research Council Centre of Excellence Scheme (EQuS CE110001013).\\

* email: david.reilly@sydney.edu.au

\bibliographystyle{apsrev4-1}
%

\end{document}